# Chemical Engineering of Altermagnetism in Two-Dimensional Metal-Organic Frameworks

*Diego López-Alcalá, Alberto M. Ruiz, Andrei Shumilin and José J. Baldoví\**

Instituto de Ciencia Molecular, Universitat de València, Catedrático José Beltrán 2, 46980 Paterna, Spain.

E-mail: j.jaime.baldovi@uv.es

**ABSTRACT**

Altermagnetism represents a novel class of collinear antiferromagnetism exhibiting non-relativistic spin splitting without net magnetization, driven by lattice symmetry rather than spin-orbit coupling (SOC). Here, we introduce a coordination-driven chemical strategy to realize altermagnetic (AM) spin splitting in two-dimensional (2D) planar tetracoordinated Cr-based metal-organic frameworks (MOFs). Using density functional theory (DFT) calculations, we demonstrate that replacing centrosymmetric pyrazine (pyz) ligands with non-centrosymmetric imidazole (imz) linkers in Cr-based MOFs reduces lattice symmetry, enabling g-wave AM spin splitting up to 65 meV. Furthermore, frontier molecular orbital engineering (FMOE) allows selective ligand spin polarization, inducing a shift to d-wave AM anisotropy in polycyclic ligand-based 2D MOFs with spin splitting up to 83.9 meV. Microscopic magnetic exchange interactions ($J$) analysis reveals that ligand-mediated interactions dominate over metal-metal coupling, stabilizing AM order in systems with radical ligands. Interestingly, we further confirm AM spin splitting in spin wave spectrum, where chiral magnon splitting is observed. Finally, we show that AM spin splitting gives rise to experimentally accessible charge to spin conversion, emerging as a linear response in d-wave and as a symmetry-allowed nonlinear effect in g-wave 2D AM MOFs. This work establishes coordination chemistry as a powerful and versatile route to symmetry control in 2D MOFs, enabling rational design of 2D molecular materials with tunable electronic and AM properties for next-generation spintronic devices.

**INTRODUCTION**

Spintronics has grown rapidly as controlling the spin degree of freedom enables key functionalities such as giant magnetoresistance, spin-transfer torque, and electrically driven spin-orbit coupling (SOC) effects.[1–3] These rely on magnetic interactions and spin alignment, classically described by ferromagnetic (FM) or antiferromagnetic (AFM) order. Beyond these conventional configurations, a variety of unconventional magnetic states have been identified, including non-collinear textures and topological spin configurations.[4,5] Among them, altermagnetism stands out as a new class of collinear antiferromagnets that display sizable non-relativistic spin splitting despite zero net magnetization.[6,7] This behavior originates from the lattice symmetry: crystallographic rotation operations relate opposite spin sublattices in a way that cannot be reproduced by inversion ($i$), translation ($t$) or their combinations.[8] Consequently, the degeneracy between opposite spin channels is lifted in momentum space while the real-space magnetic structure remains strictly AFM. This unique combination of AFM order and momentum-space spin splitting enables technologically relevant functionalities, including highly anisotropic spin currents, efficient spin filtering without SOC, and electrically controllable spin transport responses.[9] In this context, several strategies have been developed to engineer the symmetry conditions required for altermagnetism, including the application of external electric fields, Janus architectures, and twistronics.[10] Experimentally, room-temperature altermagnetic (AM) spin splitting has been confirmed both in the electronic structure by angle-resolved photoemission spectroscopy (ARPES)[11,12] and in the spin-wave spectrum through inelastic neutron scattering (INS) measurements.[13]

Metal-organic frameworks (MOFs) are periodic architectures in which metal centers are connected through organic ligands.[14–16] Their high degree of chemical tunability, rooted in the versatility of the ligands, has positioned MOFs as platforms for unprecedented functionalities across diverse fields, leading to their recognition with the Nobel Prize in Chemistry in 2025.[17] In particular, magnetic MOFs have recently emerged as promising candidates for spintronic devices.[18–21] Taking advantage from coordination chemistry, emerging novel magnetic phenomena can be induced in molecular solids.[22,23] Remarkably, Pedersen et al. demonstrated that incorporating open-shell organic linkers enables strong metal-radical magnetic interactions in layered $CrCl_2(pyz)_2$ (pyz = pyrazine),[24] where post-synthetic reduction yields robust ferrimagnetism with Curie temperatures up to 515 K and large coercivity.[25,26] Beyond bulk crystals, exfoliation down to the monolayer limit has been achieved in related pyz-based layered MOFs[27,28] and theoretically predicted to be experimentally feasible in $CrCl_2(pyz)_2$.[29,30] Building on the remarkable properties of pyz-based systems, a rapidly expanding landscape of structural and electronic modifications has been proposed to engineer new functionalities. Tunable electronic and magnetic order has been achieved through valence tautomerism,[31–33] metal substitution,[34] and applied pressure,[35] while additional phenomena such as multiferroicity[36,37] and 1D Cr–pyz motifs[38] have recently been realized. Beyond pyz linkers, planar tetracoordinated Cr-based MOFs have also been predicted to host novel functionalities and enhanced magnetic behavior through frontier molecular orbital engineering (FMOE).[39–42]

In this context, magnetic MOFs emerge as a promising platform for realizing AM lattices, since the chemical versatility of coordination networks enables the design of structures with tailored symmetrical properties. Theoretical investigations of 2D MOFs candidates for altermagnetism published to date have typically relied on symmetry breaking via selective spin polarization on the ligands,[43,44] employing bilayer stacking,[45] or constructing mathematical lattice architectures.[46] However, symmetry breaking driven by coordination chemistry remains largely unexplored in the literature, representing a promising route for designing 2D AM MOFs with tunable properties. By leveraging the vast library of organic linkers that fulfill the symmetry requirements for AM spin splitting, a large number of candidate MOFs can be engineered with tailored electronic and magnetic properties.

In this work, we demonstrate that coordination-driven symmetry breaking provides a powerful route to tailor AM spin splitting in 2D MOFs. We first identify clear evidence of AM spin splitting in a 2D MOF by replacing centrosymmetric ligands with non-centrosymmetric linkers, which break the lattice symmetry and yields a g-wave AM spin splitting. Furthermore, we employ FMOE to selectively induce spin polarization on the ligand scaffold in related non-centrosymmetric ligand based 2D MOFs. Remarkably, the sublattice symmetry breaking driven by ligand spin polarization produces a transition to d-wave AM anisotropy. Moreover, varying the degree of conjugation within the organic linkers enables systematic transitions from insulating to narrow band gap semiconducting states. Finally, we address the spin-dependent transport fingerprints of AM spin splitting. While d-wave altermagnets exhibit a spin-dependent conductivity, symmetry forbids such a response in g-wave systems, where spin splitting instead emerges as a nonlinear effect at third order in the applied electric field. We show that both regimes give rise to experimentally accessible charge to spin conversion in 2D AM MOFs. Together, these results establish a general strategy for the rational design of altermagnetism in 2D MOFs, enabling coordination chemistry to tune electronic and magnetic properties in molecular materials for spintronics applications.

## RESULTS AND DISCUSSION

**Coordination-driven design strategy and structures of 2D AM MOFs**

As illustrated in Figure 1a, the AFM coupling between metal centers in a 2D MOF lattice coordinated by pyz ligands gives rise to a conventional AFM electronic band structure. This occurs because the two spin sublattices are related through $C_2$ rotation in the spin space followed by a space operation ($S$) (where $S$ represents spatial $t$ or $i$). A common way to express these symmetry-enforced connections is the notation $[X \parallel Y]$, where the operator on the left acts exclusively on the spin space and the one on the right on the lattice space.[6] Therefore, AM spin splitting in the band structure is forbidden in M(pyz)$_2$, and both spin channels remain energetically degenerate, since $[C_2 \parallel S]$ maps one spin sublattice onto the other. This protection originates from the $D_{2h}$ point group of the centrosymmetric pyz ligand, which preserves $[C_2 \parallel S]$ throughout the lattice.

In this context, a viable strategy to induce controlled symmetry breaking in 2D MOF lattices based on M(pyz)$_2$ requires the rational chemical design of a lattice with a non-centrosymmetric analogue of the pyz ligand. Imidazole (imz) emerges as an ideal candidate for this purpose, as its molecular structure lacks an inversion center due to its $C_{2v}$ point-group symmetry.[47] Figure 1b illustrates a 2D M(imz)$_2$ lattice in which the metal centers remain AFM coupled, in close analogy to the pyz-based structure. In this case, the deformation introduced by the non-centrosymmetric ligand generates an alternating pattern in the local metal coordination, giving rise to two inequivalent real space sublattices within the 2D MOF. This structural arrangement breaks $[C_2 \parallel S]$, and as a result AM spin splitting appears in the electronic band structure. Symmetry breaking induces a non-degeneracy between the two spin components, which become anisotropic in reciprocal space. Therefore, the described low symmetry 2D MOF lattice fulfills the symmetry requirements of altermagnetism, which benefits from $[C_2 \parallel S]$ breaking while preserving [$C_2 \parallel A$] (where $A$ represents a different space symmetry operation than $S$, such as $C_4$ rotation of glide mirror (g)).

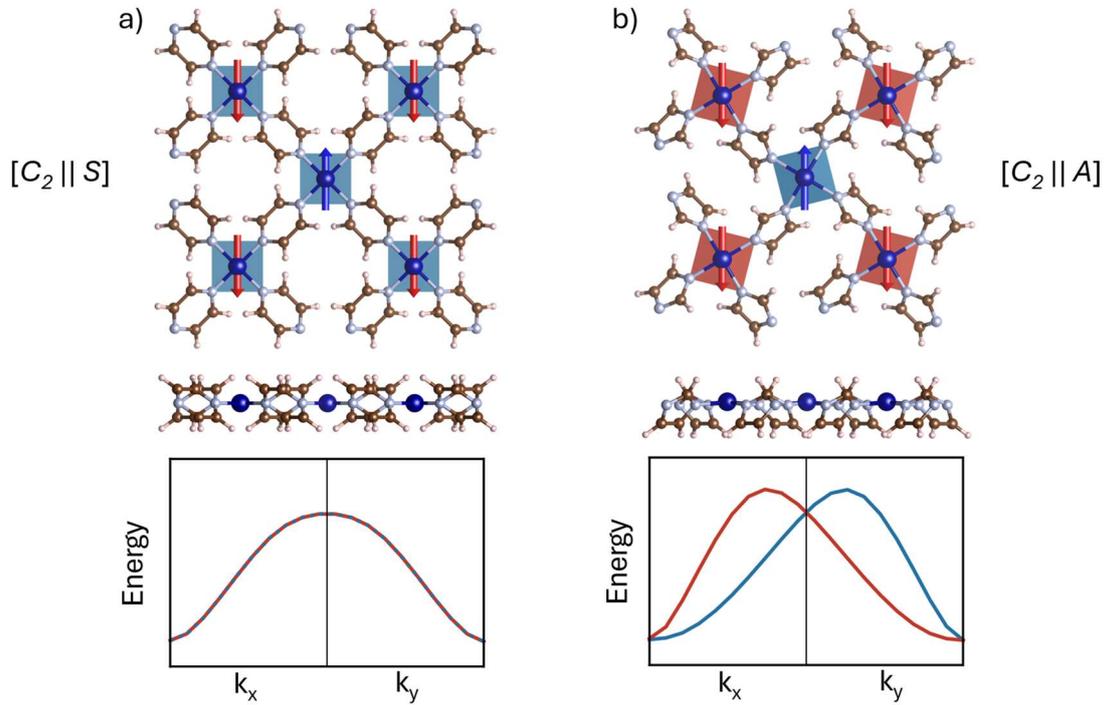

**Figure 1:** a) Representative pyz-based 2D MOF antiferromagnetically coupled, where [$C_2 \parallel S$] connects spin sublattices and gives rise to conventional AFM spin degeneracy. b) Representative imz-based 2D MOF antiferromagnetically coupled, where non-centrosymmetric ligands break [$C_2 \parallel S$] and give rise to AM spin splitting. Color code: blue (metal), brown (C), cyan (N) and white (H).

Consequently, we perform density functional theory (DFT) calculations in 2D Cr(imz)$_2$ lattice to study the origin of symmetry breaking and its implication on the origin of altermagnetism in this 2D MOF structure. We consider Cr atoms as a metal center for the formation of Cr(imz)$_2$ lattice since it has been extensively studied in related pyz-analogue MOFs. The resulting structure belongs to the *P4bm* (#100) space group with lattice parameters $a = b = 8.76$ Å. Non-centrosymmetric imz ligand generates alternating patterns of clockwise and counterclockwise orientation of the ligands around the metal centers (Figure 1b). A small deviation of ~ 5 ° from planarity in square-planar coordinated Cr atoms is observed due to the ligand environment and the tilting of the imz rings is 47.7 ° from the 2D plane. To accurately describe electronic structure of 2D MOFs we employ hybrid HSE06 functional. Figure 2a shows the electronic band structure and projected density of states (PDOS) of Cr(imz)$_2$, where an indirect band gap (X → Γ) of 4.8 eV is deduced. An absence of spin splitting along Γ – X – M – Γ high-symmetry path is observed whereas along X – Y path one can notice that different spin component bands are non-degenerated. Equally occupied states of spin up and down in PDOS can confirm the altermagnetism in Cr(imz)$_2$ since there is no net magnetization along the lattice whereas a noticeable spin splitting is present at the band structure. Figure 2b shows the AM splitting at both valence (VB) and conduction bands (CB), which increases up to 43.8 and 65.1 meV in the region close to Fermi level in VB and CB, respectively. A complete picture of AM splitting along the 2D MOF plane is depicted in Figure 2c, where a g-wave anisotropy can be deduced from the 3 nodal planes on AM splitting, *i.e.* spin-degenerate points along the Brillouin zone. Those nodes in the AM wave correspond to the Γ – X – M – Γ high-symmetry path, whereas in the X – Y path an alternated spin splitting is observed. In Cr(imz)$_2$, g-wave AM anisotropy is controlled by the combination of both [E || $C_4$] and [$C_2$|| g] symmetries, which fulfils the symmetry requirements described above due to [$C_2$ || S] breaking. Additionally, we compute the band structure of Cr(imz)$_2$ including SOC and find no noticeable changes in the electronic bands (Figure S6), thereby confirming the non-relativistic origin of the AM spin splitting.

Altermagnetism is allowed in Cr(imz)$_2$ since AFM coupling between metal centers is observed, being the FM configuration higher in energy by 8.4 meV/Cr atom. In this system, spin density is restricted to the Cr atoms since the ligands do not show significant spin polarization (Figure 2d). Calculated magnetic moments are 3.7 μ$_B$ in Cr atoms, compatible with reported Cr(II)-based planar tetracoordinated MOFs in literature.[26] We conduct phonon analysis and *ab initio* molecular dynamics (AIMD) simulations to elucidate the structural and thermodynamic stability of Cr(imz)$_2$ 2D MOF. Figure 2e shows the calculated phonon dispersion of the structure, where no evident negative frequencies are observed in phonon modes, which aligns with a prominent dynamical stability of the structure. AIMD simulations reveal that the 2D MOF structure is preserved in room temperature simulations (Figure 2f). At temperatures up to 600K the ligand rings start to rotate due to thermal activation (Figure S2), but no evident phase transition is observed.

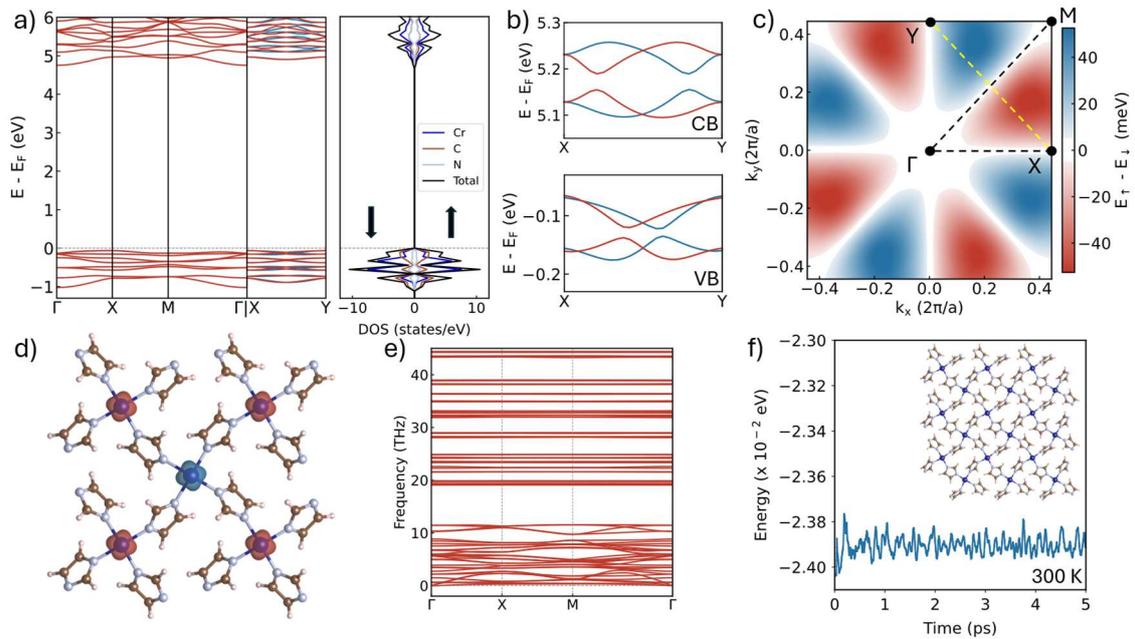

**Figure 2:** a) Band structure and PDOS of Cr(imz)$_2$ calculated using hybrid HSE06 functional. b) Zoom of CB and VB regions along high-symmetry path with AM spin splitting. c) Spin splitting in VBM along 2D MOF plane. d) Spin density in Cr(imz)$_2$. Blue (red) isosurface represents spin up (down) component. e) Phonon spectrum and f) AIMD simulation at 300 K in Cr(imz)$_2$.

**Frontier Molecular Orbital Engineering of 2D AM MOFs**

The conclusions extracted from above can be extrapolated to many 2D MOFs which fulfill the symmetry criteria for altermagnetism, due to chemical versatility of organic linkers. Therefore, we extend the electronic and magnetic properties study to the 2D MOFs depicted in Figure 3. We construct a set of Cr planar tetracoordinated lattices with non-centrosymmetric ligands analogous to imz, containing both monocyclic and polycyclic organic linkers. Analogously to Cr(imz)$_2$, a symmetry breaking between adjacent Cr sublattices is observed in all structures. Changes in ligand coordination not only alter the distance between magnetic Cr atoms in the lattice but also modify the electronic and magnetic interactions in the 2D MOFs, driven by differences in electronic distribution arising from ligand conjugation of C atoms. We compute phonon dispersions and AIMD simulations for Cr(tdz)$_2$, Cr(DApent)$_2$, and Cr(DAind)$_2$ (see Supporting Information Sections 2–4), and in all cases we observe thermodynamic stability, as no indications of phase transitions arise during the AIMD trajectories. Nevertheless, the phonon dispersions of the polycyclic-ligand MOFs, e.g. Cr(DApent)$_2$ and Cr(DAind)$_2$, display small negative frequencies (Figures S18 and S26). We attribute this to the well-known limitations of harmonic phonon calculations in flexible, low-density frameworks such as MOFs, where soft vibrational modes, anharmonicity, and large-amplitude ligand motions often lead to artificial instabilities that do not reflect the true dynamical behavior of the material.[48–50] This interpretation is consistent with the absence of structural distortions in AIMD simulations, which confirms the stability of the optimized geometries.

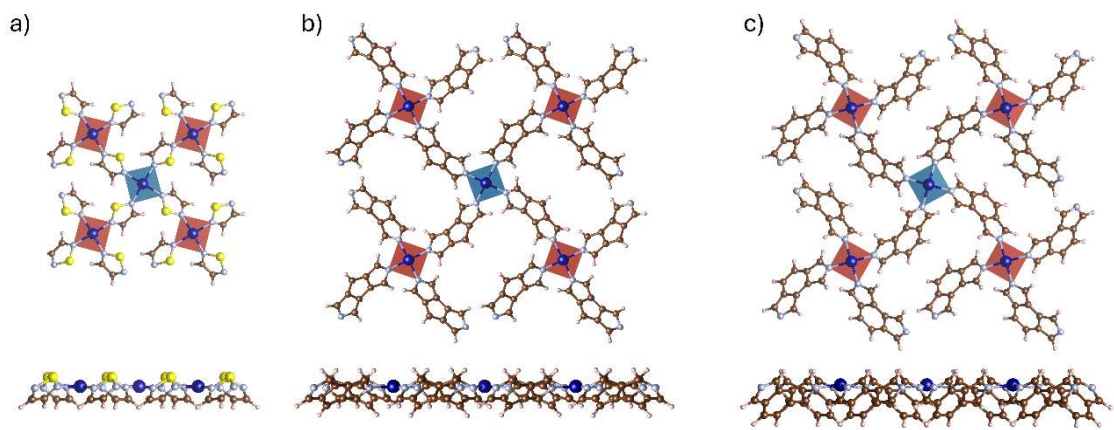

**Figure 3:** Top and side views of a) Cr(tdz)$_2$, b) Cr(DApent)$_2$ and c) Cr(DAind)$_2$. Color code: blue (metal), brown (C), cyan (N), yellow (S) and white (H).

Interestingly, we observe that the spin distribution along the lattice is strictly dependent on the energy alignment of the frontier molecular orbitals in the MOFs. Larger energy barrier between metal HOMO and ligand LUMO ($\Delta_{CT}$) leads to higher spin localization at the metal centers and prevents spin polarization at the ligands. On the other hand, reduced $\Delta_{CT}$ values promote exchange-induced spin polarization at the ligands due to a minimal energy barrier. This tautomeric switch via FMOE has been observed in similar vdW magnetic MOFs, which can lead to different magnetic ordering.[32,51] We compute $\Delta_{CT}$ for each ligand via periodic DFT calculations on each 2D MOF. Additionally, we conduct cluster DFT calculations on the isolated ligands to investigate their HOMO – LUMO gaps ($\Delta_L$) (See Supporting Information Section 5), which can be also related with the ability of the organic linker to accept spin polarization from the metal. Despite both methods are conceptually different we observed a noticeable alignment of their results, which enforces the conclusions extracted from the analysis of the impact of $\Delta_{CT}$ on the distribution of spin polarization. Figure 4a and b show the calculated $\Delta_{CT}$ and $\Delta_L$ values for each organic linker via periodic and cluster DFT, respectively. In the case of monocyclic linkers (Figure 4a), one can see that imz presents larger gaps than tdz. This aligns perfectly with the observation of confined spin polarization at the metal in Cr(imz)$_2$ and the spin polarized ligands in Cr(tdz)$_2$ (Figure S10 and S11). On the other hand, polycyclic organic linkers (Figure 4b) show the same trend for DApent and DAind, where the former shows larger gaps compared to the latter. Therefore, spin density is localized at the metal centers in Cr(DApent)$_2$ (Figure S19), whereas spin polarization is observed over the ligand scaffold in Cr(DAind)$_2$ (Figure 4c).

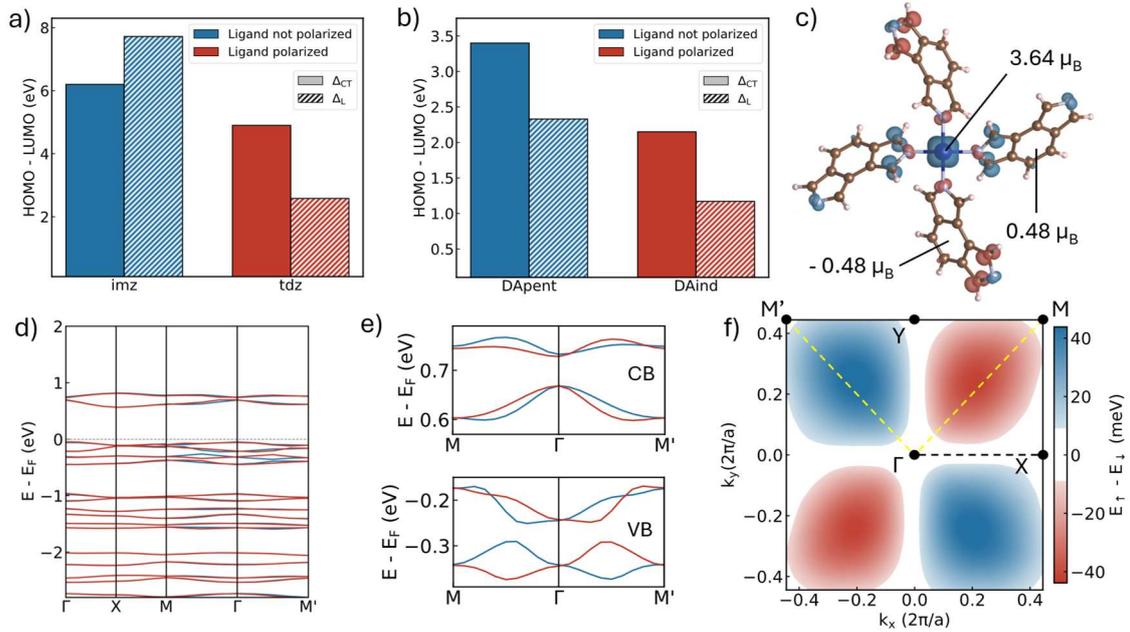

**Figure 4:** Calculated $\Delta_{CT}$ and $\Delta_L$ for a) monocyclic and b) polycyclic-based 2D MOFs. Blue (red) color represents spin unpolarized (polarized) ligands. c) Spin density with magnetic moments, d) band structure calculated using hybrid HSE06 functional, e) zoom of CB and VB regions along high-symmetry path with AM spin splitting and f) spin splitting in VBM along 2D MOF plane in Cr(DAind)$_2$.

Redistribution of spin polarization along the lattice has an enormous impact on the electronic structure of 2D MOFs. Figure 4d shows the band structure for AM Cr(DAind)$_2$, where a direct band gap ($\Gamma \rightarrow \Gamma$) of 0.61 eV is observed. This value is much smaller than the one found in Cr(imz)$_2$, which is attributed to the extended conjugation of DAind compared to imz.[52,53] This highlights the potential of rational chemical design of organic linkers, since it allows engineering MOFs with tunable band gaps from insulators to narrow band gap semiconductors. Interestingly, AM splitting in Cr(DAind)$_2$ is observed along M – Γ – M' high symmetry path (Figure 4e), which is completely different picture to the one observed in Cr(imz)$_2$. In this case we observe a d-wave anisotropy of the AM splitting, since only two nodal planes are observed at the AM splitting in the 2D lattice (Figure 4f). In this case, ligand spin polarization introduces an inequivalent set of spin-dependent moments in the lattice, breaking the symmetries compatible with g-wave anisotropy, namely [$E \parallel C_4$] and [$C_2 \parallel g$], while preserving [$C_2 \parallel g\, C_4$]. This symmetry lowering drives the transition to d-wave AM anisotropy. In Cr(DAind)$_2$, we observe an AM spin splitting of 83.9 and 35.2 meV in the region close to Fermi level in VB and CB, respectively.

**Magnetic Properties of 2D AM MOFs**

The distribution of magnetic moments across a lattice gives rise to multiple competing magnetic configurations, and a thorough understanding of their relative energies is essential to determine the magnetic ground state of a system. Accordingly, we calculate the energy difference between FM, ferrimagnetic (FiM) and AM configurations in the 2D MOFs studied in this work employing HSE06 (Table 1) and PBE+U functionals (see Supporting Information Sections 1–4). For systems with non-polarized organic linkers, such as Cr(imz)$_2$ and Cr(DApent)$_2$, a FiM configuration does not exist because the magnetic moments on the Cr atoms are identical. In both materials, the AM configuration constitutes the ground state and is separated from the competing FM state by a substantial energy gap.

**Table 1:** Relative energies of different magnetic configurations FM, FiM and AM (in meV/Cr atom), magnetic moment in Cr atoms and ligands ($M_{Cr}$ and $M_L$, in $\mu_B$), magnetic exchange interactions ($J$, in meV) and magnetic anisotropy ($D$, in $\mu$eV/Cr atom).

|        | imz    | tdz    | DApent | DAind  |
|--------|--------|--------|--------|--------|
| **FM**     | 8.4    | 290.25 | 15.14  | 157.57 |
| **FiM**    | -      | 0.00   | -      | 2.41   |
| **AM**     | 0.00   | 293.76 | 0.00   | 0.00   |
| $M_{Cr}$   | 3.71   | 3.65   | 3.65   | 3.64   |
| $M_L$      | -      | 0.56   | -      | 0.48   |
| $J_1$      | -1.85  | 1.72   | -3.20  | 1.45   |
| $J_2$      |        | -22.62 |        | 3.34   |
| $J_{2'}$   | -      | -22.62 | -      | -13.37 |
| $D$        | 406.51 | 377.57 | 376.12 | 205.9  |

A different scenario is observed in 2D MOFs with spin-polarized organic linkers. Here, magnetic moments of 0.56 and 0.48 $\mu_B$ are observed on the tdz and DAind ligands, respectively, consistent with an effective reduction of the ligand scaffold. The spin polarization on these radical ligands exhibits a strong tendency to couple antiferromagnetically with metal spins, which has been observed in similar pyz-based MOFs.[25,26] The dominant role of metal-ligand interactions over metal-metal interactions dictates that magnetic ordering is largely governed by the former. As a result, both the FiM and AM configurations are stabilized to a much greater extent than the FM configuration. Figure 5a shows the magnetic exchange interactions ($J$) between neighboring sites. Here, $J_1$ denotes the metal-metal magnetic interaction mediated by the organic linkers. When spin polarization is induced on the ligands, additional metal-ligand exchange arises along the lattice ($J_2$). This interaction is further split into $J_2$ and $J_2'$, depending on whether the radical couples ferromagnetically or antiferromagnetically with the neighboring metal, respectively. We employ this notation and the results presented in Table 1 to map a spin Hamiltonian as follows:

$$H = -\sum_{i \neq j} J_{ij} \vec{S_i} \cdot \vec{S_j} - \sum_i D_i \vec{S_i^2} \qquad (1)$$

where $J_{ij}$ represents the magnetic interaction between two magnetic moments along the lattice ($S_i$ and $S_j$) and $D$ is the magnetic anisotropy. In this expression a positive (negative) value of $J$ favors FM (AFM) coupling between spin moments.

As one can observe in Table 1, in the case of Cr(tdz)$_2$ the FiM configuration is the ground state, which is stabilized due to the dominant AFM contribution of $J_2$. Thus, no splitting is observed between $J_2$ and $J_2'$ since FiM configuration allows symmetric AFM metal-ligand interactions. A net magnetization is present in Cr(tdz)$_2$ MOF due to FiM coupling between metals and radicals, which is not compatible with altermagnetism. A different scenario is observed in Cr(DAind)$_2$, where AM configuration is the lowest energy magnetic ordering. In this case, the alternated spin-polarization over the metal-ligand magnetic structure generates symmetry breaking between $J_2$ and $J_2'$ (see Supporting Information Section 6). AFM metal-ligand coupling leads to large negative $J_2'$, which results in the stabilization of AM order in Cr(DAind)$_2$. Interestingly, $J_2$ is FM but presents lower intensity than $J_2'$. We attribute the disparity in the magnitude of $J_2$ and $J_2'$ primarily to an asymmetric spin density distribution in DAind (Figure S44) and the specific orbital ordering of the high-spin quasi-square planar Cr(II) center. Figure 5b illustrates the calculated energy alignment of the $d$ orbitals in Cr(DAind)$_2$, revealing that the $d_{x^2-y^2}$ orbital remains unoccupied, a configuration that fosters the FM interaction observed in $J_2$. This interaction through an empty orbital leads to less electronic repulsion and closer metal-ligand distance. This interpretation is corroborated by bond length analysis, which exhibits a shorter metal-ligand bond length for the FM pathway compared to the slightly larger one (~ 0.5%) associated with the AFM interaction. Additionally, we compute $J_2$ via cluster DFT calculations using a broken symmetry approach to corroborate its remarkable high intensity (see Supporting Information Section 5.5). Broken symmetry calculations align with periodic DFT calculations since we also observe high intensity magnetic interactions and enhanced $J_2$ in Cr(tdz)$_2$ compared to Cr(DAind)$_2$.

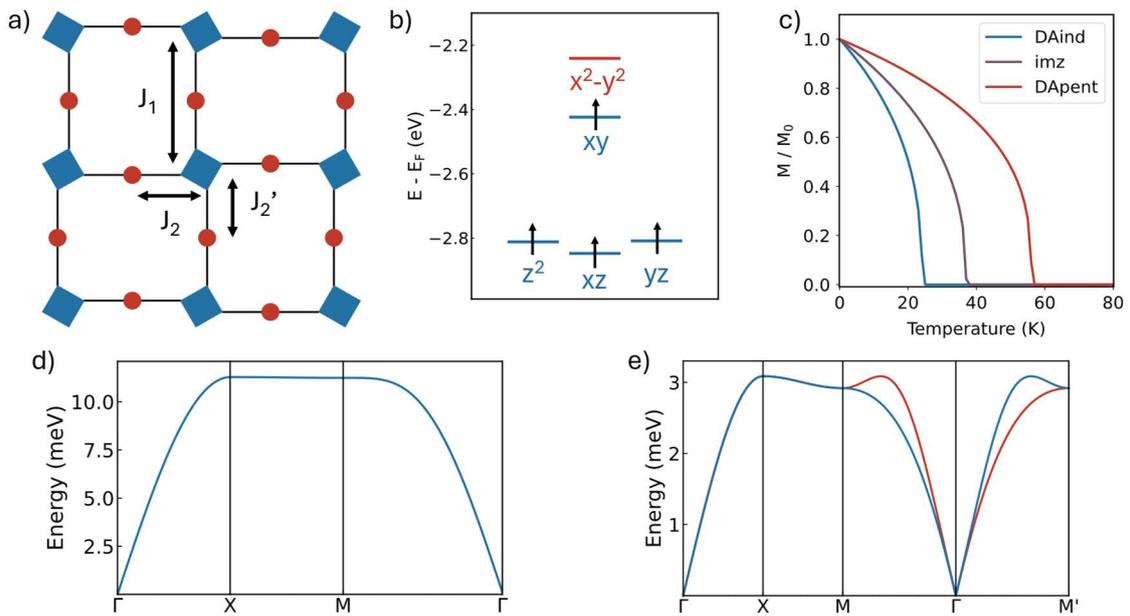

**Figure 5:** a) Schematic representation of magnetic interactions in imz-based 2D MOFs. Metal centers are represented as blue squares and ligands as red dots. b) $d$ orbital energy alignment of Cr atoms in Cr(DAind)$_2$. c) Calculated $T_N$ in Cr(imz)$_2$, Cr(DApent)$_2$ and Cr(DAind)$_2$ via atomistic spin dynamics simulations. Simulated spin-wave spectra in d) g-wave AM Cr(imz)$_2$ and e) d-wave AM Cr(DAind)$_2$. Color code: blue (red) lines represent left-handed (right-handed) magnons with negative (positive) chirality.

Next, we calculate $D$ as the energy difference between a spin aligned within the MOF plane and aligned out of the plane configurations. In all systems, the preferred orientation is out of plane, with only minor ligand-dependent variations. This trend is consistent with previous $D$ calculations on planar tetracoordinated Cr-based 2D MOFs, where an orthogonal spin orientation is also favored.[40,44] Figure 5c presents the Néel temperatures ($T_N$) obtained from atomistic spin-dynamics simulations of the representative 2D AM MOFs described in this work. Notably, the simulated $T_N$ of Cr(DApent)$_2$ lies close to liquid nitrogen temperature (77 K), reinforcing its potential for AM-based cryogenic spintronic applications. In contrast, the calculated $T_N$ values for Cr(DAind)$_2$ and Cr(imz)$_2$ are significantly lower, mainly due to the competing FM $J_1$ in Cr(DAind)$_2$ and the limited strength of magnetic interactions in Cr(imz)$_2$. Spin wave dispersions in both d-wave and g-wave AM MOFs show linear dispersion near Γ-point, typical for AFM materials (Figure 5d and e). Interestingly, magnon dispersion in Cr(DApent)$_2$ displays the chirality-based spin splitting consistent with the d-wave AM anisotropy. On the other hand, in the case of g-wave Cr(imz)$_2$, the spin-Hamiltonian is controlled by the nearest neighbor metal-metal exchange interactions, which increases the effective symmetry and results in degenerate magnon modes (See Supporting Information Section 7).

**Spin-dependent transport simulations**

Spin splitting effect in AM materials is a non-relativistic analogue of the conventional spin Hall effect in which an applied electric field separates spin-up and spin-down charge currents.[54] In d-wave AM anisotropy, this effect can be captured by a spin-dependent conductivity tensor. On the other hand, in g-wave AM symmetry forbids distinct conductivity tensors for different spin component electrons in the linear regime. However, it was recently shown that a spin-splitting effect can nevertheless emerge as a nonlinear response, specifically in the current contribution proportional to the cube of the electric field.[55] To investigate this behavior, we compute the nonlinear transport response within the Boltzmann formalism, including terms up to third order in the electric field (see Supporting Information Section 8). Figure 6a shows the off-diagonal component of the conductivity tensor in d-wave Cr(DAind)$_2$ as a function of Fermi energy, evaluated at 300 K and a relaxation time of 10 fs. When the Fermi energy lies in either the conduction or valence band, $\sigma_{yx}$ acquires opposite signs for spin-up and spin-down electrons, resulting in non-relativistic spin-splitting angles of up to 0.4 degrees. Figure 6b displays the $\sigma_{yxxx}$ component of the nonlinear conductivity tensor in g-wave Cr(imz)$_2$, which gives rise to nonlinear spin splitting.

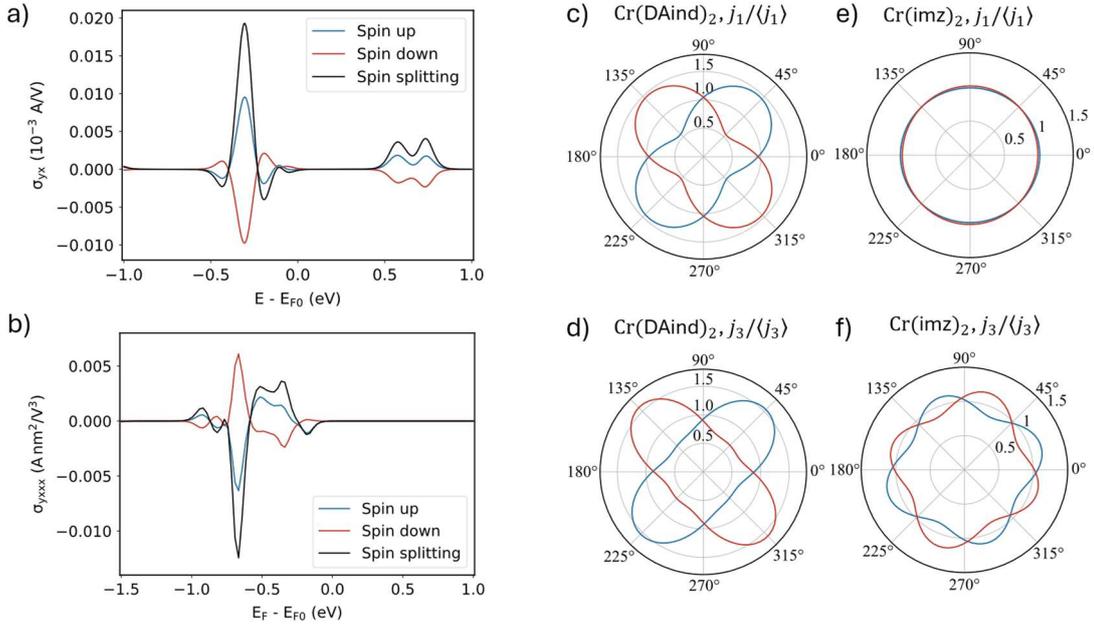

Figure 6: a) Linear spin-splitting effect in Cr(DAind)$_2$: difference between the off-diagonal components $\sigma_{yx}$ of the conductivity tensor for spin-up and spin-down electrons as a function of the Fermi energy. b) Third-order spin-splitting effect in Cr(imz)$_2$: difference between the $\sigma_{yxxx}$ components of the nonlinear conductivity tensor for spin-up and spin-down electrons as a function of the Fermi energy. (c–f) Dependence of the current density along the direction of the applied electric field on the field polar angle. Results are shown for Cr(DAind)$_2$ at $E_F - E_F(0) = -0.31$eV (c,d) and for Cr(imz)$_2$ at $E_F - E_F(0) = -0.67$eV (e,f). Panels c) and e) show the linear contribution $j_1$, while panels d) and f) show the cubic contribution $j_3$. Angle brackets denote averaging over the polar angle.

To further analyze spin-dependent transport, Figures 5c–e show the polar-angle dependence of the spin-up and spin-down currents measured along the direction of the applied field at the Fermi level positions corresponding to the strongest spin-splitting (-0.31eV for Cr(DApent)$_2$ and -0.67 eV for Cr(imz)$_2$). Finally, we separately plot the first-order and third-order contributions in the electric field. In Cr(DAind)$_2$, both contributions follow the expected d-wave symmetry: the conductivity for spin-up electrons is higher along the ΓM direction, whereas for spin-down electrons it is higher along the ΓM′ direction. For Cr(imz)$_2$, the linear-response transport is spin independent, but the third-order contribution exhibits the characteristic g-wave symmetry, as shown in Figure 5f.

**CONCLUSIONS**

In summary, we present a general design strategy for engineering AM non-relativistic spin splitting in 2D MOFs through coordination-driven lattice symmetry breaking. We show that employing non-centrosymmetric ligands directly introduces an inequivalence in the square planar Cr sublattices, generating a sizable g-wave AM splitting of 65.0 meV in Cr(imz)$_2$. This represents the first demonstration that coordination chemistry alone can be used to tailor altermagnetism in low dimensional molecular materials. Leveraging the broad chemical space of organic ligands, we further construct related 2D MOFs where increased π-conjugation enables systematic control over both electronic structure and magnetic behavior. We then apply FMOE to selectively induce spin polarization on the ligand scaffold in imz-derived MOFs. The emergence of ligand-localized magnetic moments breaks the sublattice equivalence and drives a transition to d-wave AM anisotropy, reaching spin splitting up to 83.9 meV in Cr(DAind)$_2$. In these systems, strong metal-ligand interactions stabilize the magnetic ground state and preserve the AM phase. Interestingly, we show that AM spin splitting manifests not only in the electronic band structure but also in the spin-wave spectrum, where chiral magnon splitting is observed in d-wave Cr(DAind)$_2$. Finally, we show that AM spin splitting directly manifests in charge transport as a non-relativistic analogue of the spin Hall effect. In d-wave Cr(DAind)$_2$, spin-dependent linear conductivity leads to sizable spin-splitting angles, while in g-wave Cr(imz)$_2$ symmetry forbids linear spin separation but allows a distinct nonlinear response emerging at third order in the electric field. Importantly, both regimes generate measurable spin-polarized currents under experimentally realistic conditions, making nonlinear transport a viable probe of altermagnetism in 2D MOF monolayers. Together, these results demonstrate that coordination chemistry provides a powerful and versatile route to design 2D MOFs with tunable AM properties, opening new opportunities for chemically engineered spin functionalities in coordination solids.

**METHODS**

Structural relaxations, electronic structure, magnetic configurations, AIMD and phonon calculations are performed using VASP package.[56] We employ the generalized gradient approximation (GGA) to describe the exchange-correlation energy, in combination with PBE functional. Given the close interaction between adjacent ligands, we employ Grimme D3 vdW corrections. We introduce a vacuum layer of 15 Å to avoid periodic interactions between 2D layers. To accurately describe the partially filled *3d* orbitals of Cr in GGA-PBE calculations we use Hubbard U correction (See Supporting Information for DFT+U results). To accurately describe the electronic and magnetic properties we calculate the electronic band structures and the relative energy of each magnetic configuration using hybrid HSE06 functional (See Supporting Information for comparison with DFT+U results). The projector augmented-wave (PAW) pseudopotential in combination with a plane-wave cutoff of 500 eV is used. A Monkhorst-Pack k-point mesh of 3x3x1 (2x2x1) is used for monocyclic (polycyclic)-based 2D MOFs. We perform geometrical optimization with a convergence criterion of 0.01 eV/A. AIMD simulations are performed for 5 ps with a time step of 1 fs at 300K and 600K using the canonical NVT ensemble. Phonon dispersions are calculated using Phonopy code.[57] Magnetic parameters demanded to map the spin Hamiltonian (Equation 1) are calculated using SIESTA software,[58,59] in order to take benefit from its localized atomic orbital approach. We use PBE+U ($U_{eff}$ = 3 eV) in combination with double-$\zeta$ polarized basis set and a real-space mesh cutoff of 900 Ry. SOC is considered for $D$ calculations, where we analyze the energy difference between in plane and out of plane spin orientations. Magnetic interactions ($J$) are calculated using the interface between SIESTA and TB2J code.[60] Cluster DFT calculations are performed using Gaussian09 software in its revision D01,[61] where we employ B3LYP functional in combination with Def2TZVP basis set. Maximally localized Wannier functions are constructed using Wannier90,[62] with the $d$ orbitals of Cr, the $s$ and $p$ orbitals of C, the $p$ orbitals of N and the $s$ orbitals of H as the basis, to generate a tight-binding Hamiltonian for spin-dependent transport calculations. Transport calculations are made within Boltzmann formalism with a single phenomenological relaxation time ($\tau$) of 10 fs and a temperature of 300 K.

AUTHOR INFORMATION

Corresponding author

*E-mail: j.jaime.baldovi@uv.es

Author contributions

The manuscript was written through contributions of all authors. This work is part of the PhD thesis of D.L-A. All authors have given approval to the final version of the manuscript.

Notes

The authors declare no competing financial interest.

ACKNOWLEDGEMENTS

The authors acknowledge financial support from the European Union (ERC-2021-StG101042680 2D-SMARTiES), the Spanish Government MCIU (PID2024-162182NA-I00 2D-MAGIC) and the Generalitat Valenciana (grant CIDEXG/2023/1). A.M.R. thanks the Spanish MIU (Grant No FPU21/04195). The calculations were performed on the HAWK cluster of the 2D Smart Materials Lab hosted by Servei d'Informàtica of the Universitat de València.

REFERENCES

(1) Žutić, I.; Fabian, J.; Das Sarma, S. Spintronics: Fundamentals and Applications. *Rev Mod. Phys.* 2004, *76*, 323–410.
(2) Sierra, J. F.; Fabian, J.; Kawakami, R. K.; Roche, S.; Valenzuela, S. O. Van Der Waals Heterostructures for Spintronics and Opto-Spintronics. *Nat. Nanotechnol.* 2021, *16*, 856–868.
(3) Manchon, A.; Železný, J.; Miron, I. M.; Jungwirth, T.; Sinova, J.; Thiaville, A.; Garello, K.; Gambardella, P. Current-Induced Spin-Orbit Torques in Ferromagnetic and Antiferromagnetic Systems. *Rev. Mod. Phys.* 2019, *91* (3), 035004.
(4) Zhou, Y.; Li, S.; Liang, X.; Zhou, Y. Topological Spin Textures: Basic Physics and Devices. *Advanced Materials* 2025, *37*, 2312935.
(5) Rimmler, B. H.; Pal, B.; Parkin, S. S. P. Non-Collinear Antiferromagnetic Spintronics. *Nat. Rev. Mater.* 2024, *10*, 109–127.
(6) Šmejkal, L.; Sinova, J.; Jungwirth, T. Beyond Conventional Ferromagnetism and Antiferromagnetism: A Phase with Nonrelativistic Spin and Crystal Rotation Symmetry. *Phys. Rev. X* 2022, *12*, 031042.
(7) Song, C.; Bai, H.; Zhou, Z.; Han, L.; Reichlova, H.; Dil, J. H.; Liu, J.; Chen, X.; Pan, F. Altermagnets as a New Class of Functional Materials. *Nat. Rev. Mater.* 2025, *10*, 473–485.
(8) Fender, S. S.; Gonzalez, O.; Bediako, D. K. Altermagnetism: A Chemical Perspective. *J Am. Chem. Soc.* 2025, *147*, 2257–2274.
(9) Šmejkal, L.; Sinova, J.; Jungwirth, T. Emerging Research Landscape of Altermagnetism. *Phys. Rev. X* 2022, *12*, 040501.
(10) Bai, L.; Feng, W.; Liu, S.; Šmejkal, L.; Mokrousov, Y.; Yao, Y. Altermagnetism: Exploring New Frontiers in Magnetism and Spintronics. *Adv. Funct. Mater.* 2024, *34*, 2409327.
(11) Krempaský, J.; Šmejkal, L.; D'Souza, S. W.; Hajlaoui, M.; Springholz, G.; Uhlířová, K.; Alarab, F.; Constantinou, P. C.; Strocov, V.; Usanov, D.; Pudelko, W. R.; González-Hernández, R.; Birk Hellenes, A.; Jansa, Z.; Reichlová, H.; Šobáň, Z.; Gonzalez Betancourt, R. D.; Wadley, P.; Sinova, J.; Kriegner, D.; Minár, J.; Dil, J. H.; Jungwirth, T. Altermagnetic Lifting of Kramers Spin Degeneracy. *Nature* 2024, *626*, 517–522.
(12) Jiang, B.; Hu, M.; Bai, J.; Song, Z.; Mu, C.; Qu, G.; Li, W.; Zhu, W.; Pi, H.; Wei, Z.; Sun, Y.-J.; Huang, Y.; Zheng, X.; Peng, Y.; He, L.; Li, S.; Luo, J.; Li, Z.; Chen, G.; Li, H.; Weng, H.; Qian, T. A Metallic Room-Temperature d-Wave Altermagnet. *Nat. Phys.* 2025, *21* (5), 754–759.
(13) Liu, Z.; Ozeki, M.; Asai, S.; Itoh, S.; Masuda, T. Chiral Split Magnon in Altermagnetic MnTe. *Phys. Rev. Lett.* 2024, *133*, 156702.
(14) James, S. L. Metal-Organic Frameworks. *Chem. Soc. Rev.* 2003, *32*, 276.
(15) Zhou, H.-C.; Long, J. R.; Yaghi, O. M. Introduction to Metal–Organic Frameworks. *Chem. Rev.* 2012, *112*, 673–674.

(16) Zhou, H.-C. "Joe"; Kitagawa, S. Metal–Organic Frameworks (MOFs). *Chem. Soc. Rev.* 2014, *43*, 5415–5418.
(17) Lee, J.-S. M. Medal for Metal–Organic Frameworks. *Nature Synthesis* 2025, *4*, 1485–1485.
(18) Coronado, E.; Mínguez Espallargas, G. Dynamic Magnetic MOFs. *Chem. Soc. Rev.* 2013, *42*, 1525–1539.
(19) Coronado, E. Molecular Magnetism: From Chemical Design to Spin Control in Molecules, Materials and Devices. *Nat. Rev. Mater.* 2019, *5*, 87–104.
(20) Thorarinsdottir, A. E.; Harris, T. D. Metal–Organic Framework Magnets. *Chem Rev* 2020, *120* (16), 8716–8789. https://doi.org/10.1021/acs.chemrev.9b00666.
(21) Yan, X.; Su, X.; Chen, J.; Jin, C.; Chen, L. Two-Dimensional Metal-Organic Frameworks Towards Spintronics. *Angewandte Chemie* 2023, *135*, e202305408.
(22) Pitcairn, J.; Iliceto, A.; Cañadillas-Delgado, L.; Fabelo, O.; Liu, C.; Balz, C.; Weilhard, A.; Argent, S. P.; Morris, A. J.; Cliffe, M. J. Low-Dimensional Metal–Organic Magnets as a Route toward the $S = 2$ Haldane Phase. *J. Am. Chem. Soc.* 2023, *145*, 1783–1792.
(23) Pitcairn, J.; Ongkiko, M. A.; Iliceto, A.; Speakman, P. J.; Calder, S.; Cochran, M. J.; Paddison, J. A. M.; Liu, C.; Argent, S. P.; Morris, A. J.; Cliffe, M. J. Controlling Noncollinear Ferromagnetism in van Der Waals Metal–Organic Magnets. *J. Am. Chem. Soc.* 2024, *146*, 19146–19159.
(24) Pedersen, K. S.; Perlepe, P.; Aubrey, M. L.; Woodruff, D. N.; Reyes-Lillo, S. E.; Reinholdt, A.; Voigt, L.; Li, Z.; Borup, K.; Rouzières, M.; Samohvalov, D.; Wilhelm, F.; Rogalev, A.; Neaton, J. B.; Long, J. R.; Clérac, R. Formation of the Layered Conductive Magnet CrCl2(Pyrazine)2 through Redox-Active Coordination Chemistry. *Nat. Chem.* 2018, *10*, 1056–1061.
(25) Perlepe, P.; Oyarzabal, I.; Mailman, A.; Yquel, M.; Platunov, M.; Dovgaliuk, I.; Rouzières, M.; Négrier, P.; Mondieig, D.; Suturina, E. A.; Dourges, M.-A.; Bonhommeau, S.; Musgrave, R. A.; Pedersen, K. S.; Chernyshov, D.; Wilhelm, F.; Rogalev, A.; Mathonière, C.; Clérac, R. Metal-Organic Magnets with Large Coercivity and Ordering Temperatures up to 242°C. *Science (1979)* 2020, *370*, 587–592.
(26) Huang, Y.; Zhang, Q.; Li, Y. C.; Yao, Y.; Hu, Y.; Ren, S. Chemical Tuning Meets 2D Molecular Magnets. *Advanced Materials* 2023, *35*, 2208919.
(27) Hermosa, C.; Horrocks, B. R.; Martínez, J. I.; Liscio, F.; Gómez-Herrero, J.; Zamora, F. Mechanical and Optical Properties of Ultralarge Flakes of a Metal–Organic Framework with Molecular Thickness. *Chem. Sci.* 2015, *6*, 2553–2558.
(28) McKenzie, J.; Pennington, D. L.; Ericson, T.; Cope, E.; Kaufman, A. J.; Cozzolino, A. F.; Johnson, D. C.; Kadota, K.; Hendon, C. H.; Brozek, C. K. Tunable Interlayer Interactions in Exfoliated 2D van Der Waals Framework Fe(SCN) 2 (Pyrazine) 2. *Advanced Materials* 2024, *36*, 2409959.
(29) Li, X.; Lv, H.; Liu, X.; Jin, T.; Wu, X.; Li, X.; Yang, J. Two-Dimensional Bipolar Magnetic Semiconductors with High Curie-Temperature and Electrically Controllable Spin Polarization Realized in Exfoliated Cr(Pyrazine)2 Monolayers. *Sci. China. Chem.* 2021, *64*, 2212–2217.
(30) López-Alcalá, D.; Ruiz, A. M.; Baldoví, J. J. Exploring Spin-Phonon Coupling in Magnetic 2D Metal-Organic Frameworks. *Nanomaterials* 2023, *13*, 1172.
(31) Dunstan, M. A.; Pedersen, K. S. Valence Tautomerism, Non-Innocence, and Emergent Magnetic Phenomena in Lanthanide-Organic Tessellations. *Chemical Communications* 2025, *61*, 627–638.
(32) Aribot, F.; Voigt, L.; Dunstan, M. A.; Wan, W.; McPherson, J. N.; Kubus, M.; Yutronkie, N. J.; McMonagle, C. J.; Coletta, M.; Manvell, A. S.; Perlepe, P.; Viborg, A.; Wong, S.; Stampe, K. A.; Baran, V.; Senyshyn, A.; Deylamani, S. T.; Le, M. D.; Walker, H. C.; Chanda, A.; Trier, F.; Pryds, N.; Wilhelm, F.; Jinschek, J. R.; Pinkowicz, D.; Probert, M. R.; Clérac, R.; Christensen, N. B.; Brechin, E. K.; Rogalev, A.; Pedersen, K. S. Molecular Alloying Drives Valence Change in a van Der Waals Antiferromagnet. *Chem* 2025, *11*, 102557.

(33) Dunstan, M. A.; Manvell, A. S.; Yutronkie, N. J.; Aribot, F.; Bendix, J.; Rogalev, A.; Pedersen, K. S. Tunable Valence Tautomerism in Lanthanide–Organic Alloys. *Nat. Chem.* 2024, *16*, 735–740.

(34) Perlepe, P.; Oyarzabal, I.; Voigt, L.; Kubus, M.; Woodruff, D. N.; Reyes-Lillo, S. E.; Aubrey, M. L.; Négrier, P.; Rouzières, M.; Wilhelm, F.; Rogalev, A.; Neaton, J. B.; Long, J. R.; Mathonière, C.; Vignolle, B.; Pedersen, K. S.; Clérac, R. From an Antiferromagnetic Insulator to a Strongly Correlated Metal in Square-Lattice MCl2(Pyrazine)2 Coordination Solids. *Nat. Commun.* 2022, *13*, 5766.

(35) Huang, Y.; Pathak, A. K.; Tsai, J.-Y.; Rumsey, C.; Ivill, M.; Kramer, N.; Hu, Y.; Trebbin, M.; Yan, Q.; Ren, S. Pressure-Controlled Magnetism in 2D Molecular Layers. *Nat. Commun.* 2023, *14*, 3186.

(36) Yang, Y.; Ji, J.; Feng, J.; Chen, S.; Bellaiche, L.; Xiang, H. Two-Dimensional Organic–Inorganic Room-Temperature Multiferroics. *J .Am. Chem. Soc.* 2022, *144*, 14907–14914.

(37) Gong, J.; Sun, W.; Wu, Y.; Guo, Z.; Qian, S.; Wang, X.; Zhang, G. Ferroelectric Chirality-Driven Direction-Tunable and Spin-Invertible Corner States in 2D MOF-Based Magnetic Second-Order Topological Insulators. *Adv. Funct. Mater.* 2025, 2505270.

(38) Lou, D.; Yutronkie, N. J.; Oyarzabal, I.; Barry, H.; Mailman, A.; Dechambenoit, P.; Rouzières, M.; Wilhelm, F.; Rogalev, A.; Mathonière, C.; Clérac, R. Self-Assembly of Chromium(II) Metal-Ion and Pyrazine Ligand into One-Dimensional Coordination Polymers. *Cryst. Growth. Des.* 2025, *25*, 3996–4005.

(39) Li, X.; Yang, J. Realizing Two-Dimensional Magnetic Semiconductors with Enhanced Curie Temperature by Antiaromatic Ring Based Organometallic Frameworks. *J. Am. Chem. Soc.* 2019, *141*, 109–112.

(40) Lv, H.; Wu, D.; Cui, X.; Wu, X.; Yang, J. Enhancing Magnetic Ordering in Two-Dimensional Metal–Organic Frameworks via Frontier Molecular Orbital Engineering. *J. Phys. Chem. Lett.* 2024, *15*, 9960–9967.

(41) Li, X.; Liu, Q.-B.; Tang, Y.; Li, W.; Ding, N.; Liu, Z.; Fu, H.-H.; Dong, S.; Li, X.; Yang, J. Quintuple Function Integration in Two-Dimensional Cr(II) Five-Membered Heterocyclic Metal Organic Frameworks via Tuning Ligand Spin and Lattice Symmetry. *J. Am. Chem. Soc.* 2023, *145*, 7869–7878.

(42) Lv, H.; Li, X.; Wu, D.; Liu, Y.; Li, X.; Wu, X.; Yang, J. Enhanced Curie Temperature of Two-Dimensional Cr(II) Aromatic Heterocyclic Metal–Organic Framework Magnets via Strengthened Orbital Hybridization. *Nano. Lett.* 2022, *22*, 1573–1579.

(43) Che, Y.; Lv, H.; Wu, X.; Yang, J. Realizing Altermagnetism in Two-Dimensional Metal–Organic Framework Semiconductors with Electric-Field-Controlled Anisotropic Spin Current. *Chem. Sci.* 2024, *15*, 13853–13863.

(44) Che, Y.; Chen, Y.; Liu, X.; Lv, H.; Wu, X.; Yang, J. Inverse Design of 2D Altermagnetic Metal–Organic Framework Monolayers from Hückel Theory of Nonbonding Molecular Orbitals. *JACS Au* 2025, *5*, 381–387.

(45) Che, Y.; Lv, H.; Wu, X.; Yang, J. Bilayer Metal–Organic Framework Altermagnets with Electrically Tunable Spin-Split Valleys. *J. Am. Chem. Soc.* 2025, *147*, 14806–14814.

(46) Che, Y.; Lv, H.; Wu, X.; Yang, J. Engineering Altermagnetic States in Two-Dimensional Square Tessellations. *Phys. Rev. Lett.* 2025, *135*, 036701.

(47) Chen, S.-S. The Roles of Imidazole Ligands in Coordination Supramolecular Systems. *Cryst. Eng. Comm.* 2016, *18*, 6543–6565.

(48) Bristow, J. K.; Skelton, J. M.; Svane, K. L.; Walsh, A.; Gale, J. D. A General Forcefield for Accurate Phonon Properties of Metal–Organic Frameworks. *Physical Chemistry Chemical Physics* 2016, *18*, 29316–29329.

(49) Kamencek, T.; Bedoya-Martínez, N.; Zojer, E. Understanding Phonon Properties in Isoreticular Metal-Organic Frameworks from First Principles. *Phys. Rev. Mater.* 2019, *3*, 116003.

(50) Tan, J.-C. Fundamentals of MOF Mechanics & Structure–Mechanical Property Relationships. In *Mechanical Behaviour of Metal – Organic Framework Materials*; The Royal Society of Chemistry, 2023; pp 1–64. DOI: 10.1039/9781839166594-00001.

(51) Pitcairn, J.; Ongkiko, M. A. T.; Speakman, P. J.; Tidey, J. P.; Jordan, J.; Newton, G. N.; Stewart, J. R.; Manuel, P.; Morris, A. J.; Cliffe, M. J. Enhancing Superexchange through Frontier Orbital Engineering in a van Der Waals Metal-Organic Magnet. *ChemRxiv* 2025, DOI: 10.26434/chemrxiv-2025-9z07g.

(52) Beaujuge, P. M.; Reynolds, J. R. Color Control in π-Conjugated Organic Polymers for Use in Electrochromic Devices. *Chem. Rev.* 2010, *110*, 268–320.

(53) Sheberla, D.; Bachman, J. C.; Elias, J. S.; Sun, C.-J.; Shao-Horn, Y.; Dincă, M. Conductive MOF Electrodes for Stable Supercapacitors with High Areal Capacitance. *Nat. Mater.* 2017, *16*, 220–224.

(54) González-Hernández, R.; Šmejkal, L.; Výborný, K.; Yahagi, Y.; Sinova, J.; Jungwirth, T.; Železný, J. Efficient Electrical Spin Splitter Based on Nonrelativistic Collinear Antiferromagnetism. *Phys. Rev. Lett.* 2021, *126*, 127701.

(55) Ezawa, M. Third-Order and Fifth-Order Nonlinear Spin-Current Generation in g-Wave and i-Wave Altermagnets and Perfectly Nonreciprocal Spin Current in f-Wave Magnets. *Phys. Rev. B* 2025, *111*, 125420.

(56) Kresse, G.; Furthmüller, J. Efficiency of Ab-Initio Total Energy Calculations for Metals and Semiconductors Using a Plane-Wave Basis Set. *Comput. Mater. Sci.* 1996, *6*, 15–50.

(57) Togo, A.; Chaput, L.; Tadano, T.; Tanaka, I. Implementation Strategies in Phonopy and Phono3py. *Journal of Physics: Condensed Matter* 2023, *35*, 353001.

(58) Soler, J. M.; Artacho, E.; Gale, J. D.; García, A.; Junquera, J.; Ordejón, P.; Sánchez-Portal, D. The SIESTA Method for *Ab Initio* Order-*N* Materials Simulation. *Journal of Physics: Condensed Matter* 2002, *14*, 2745–2779.

(59) García, A.; Papior, N.; Akhtar, A.; Artacho, E.; Blum, V.; Bosoni, E.; Brandimarte, P.; Brandbyge, M.; Cerdá, J. I.; Corsetti, F.; Cuadrado, R.; Dikan, V.; Ferrer, J.; Gale, J.; García-Fernández, P.; García-Suárez, V. M.; García, S.; Huhs, G.; Illera, S.; Korytár, R.; Koval, P.; Lebedeva, I.; Lin, L.; López-Tarifa, P.; Mayo, S. G.; Mohr, S.; Ordejón, P.; Postnikov, A.; Pouillon, Y.; Pruneda, M.; Robles, R.; Sánchez-Portal, D.; Soler, J. M.; Ullah, R.; Yu, V. W.; Junquera, J. Siesta: Recent Developments and Applications. *J Chem Phys* 2020, *152*, 204108.

(60) He, X.; Helbig, N.; Verstraete, M. J.; Bousquet, E. TB2J: A Python Package for Computing Magnetic Interaction Parameters. *Comput. Phys. Commun.* 2021, *264*, 107938.

(61) Frisch, M. J. Gaussian09, Revision D01. 2009.

(62) Mostofi, A. A.; Yates, J. R.; Lee, Y.-S.; Souza, I.; Vanderbilt, D.; Marzari, N. Wannier90: A Tool for Obtaining Maximally-Localised Wannier Functions. *Comput. Phys. Commun.* 2008, *178*, 685–699.